\documentclass[aps,reprint,superscriptaddress,nofootinbib,showpacs,amsmath,amssymb]{revtex4-1}
\usepackage{latexsym,graphicx,slashed,bm,color,dcolumn}
\usepackage{hyperref}

\newcommand{\be}[1]{\begin{equation}\label{#1}}
\newcommand{\beq}{\begin{equation}}
\newcommand{\eeq}{\end{equation}}
\def\ee{\end{equation}}
\newcommand{\beqn}[1]{\begin{eqnarray}\label{#1}}
\newcommand{\eeqn}{\end{eqnarray}}

\newcommand{\mat}[4]{\left(\begin{array}{cc}{#1}&{#2}\\{#3}&{#4}
\end{array}\right)}

\renewcommand{\to}{\rightarrow}

\def\ov{\overline}

\def\lsim{\raise0.3ex\hbox{$\;<$\kern-0.75em\raise-1.1ex
\hbox{$\sim\;$}}}
\def\gsim{\raise0.3ex\hbox{$\;>$\kern-0.75em\raise-1.1ex
\hbox{$\sim\;$}}}
\def\cal{\mathcal}

\def\cL{{\cal L}}
\def\cM{{\cal M}}

\def\eps{\varepsilon}

\def\rB{{\rm B}}
\def\rL{{\rm L}}

\begin{document}

\title{Antistars or antimatter cores in mirror neutron stars?  }

\author{Zurab~Berezhiani}
\email{E-mail: zurab.berezhiani@lngs.infn.it}
\affiliation{Dipartimento di Fisica e Chimica, Universit\`a di L'Aquila, 67100 Coppito, L'Aquila, Italy} 
\affiliation{INFN, Laboratori Nazionali del Gran Sasso, 67010 Assergi,  L'Aquila, Italy}





\begin{abstract} 
The oscillation of  the neutron $n$ into mirror neutron $n'$,  its partner from  dark mirror sector,  
can gradually transform an ordinary neutron star into a mixed star consisting in part of mirror dark matter.  
The implications of the reverse process taking place in the mirror neutron stars  
depend on the sign of baryon asymmetry in mirror sector.  
Namely, if it is negative, as predicted by certain baryogenesis scenarios, then $\ov{n'}-\ov{n}$ transitions 
create a core of our antimatter gravitationally trapped in the mirror star interior.   
The annihilation of accreted gas on such antimatter cores could explain the origin 
$\gamma$-source candidates,  with unusual spectrum compatible to baryon-antibaryon annihilation, 
recently identified  in the Fermi LAT catalog,   
In addition, some part of this antimatter escaping after the mergers of mirror neutron stars 
can produce the flux of cosmic antihelium and also heavier antinuclei which are 
hunted in the AMS-02 experiment.

\end{abstract}

\maketitle



 
\noindent
{\bf 1.}  
There may exist a hidden sector of particles which are the mirror (M) replicas of 
the ordinary (O) particles. Mirror parity, 
a discrete symmetry under the specular exchange of the O and M species 
(fermions, gauge  bosons and Higgs fields of two sectors),  
implies that the two worlds should have identical microphysics. In this way, all O particles: 
electron $e$, proton $p$, neutron $n$, neutrinos $\nu$, photon $\gamma$ etc. 
must have the mass-degenerate  M  twins $e'$, $p'$, $n'$, $\nu'$, $\gamma'$ etc. 
which are sterile to the interactions of our Standard Model (SM)   
but have their own SM$'$ interactions of exactly the same pattern. 
Like O matter,  during cosmological evolution the M matter 
should form nuclei, atoms, stars and planets    
which, being invisible for us in terms of O photons, can represent  
dark matter (DM) in the Universe (for reviews see e.g. \cite{Alice,Foot}).

 A specific feature of this  scenario is that any neutral  particle   
(elementary as e.g. neutrinos or composite as e.g. the neutron) 
can have maximal mixing with its M partner, 
and such twin species can effectively oscillate between each other. 
 Namely,  the active--sterile  $\nu-\nu'$ oscillations between the O and M neutrinos, 
violating the lepton numbers $\rL$ and $\rL'$ of both sectors by one unit, 
can be experimentally observed as the deficit of neutrinos.   
Analogously, active--sterile oscillation between the neutron $n$ and M neutron $n'$, 
 induced by the mass mixing 
 \be{n-npr}
\eps \, \ov{n'} n + \text{h.c.}
\ee 
violates the baryon numbers $\rB$ and $\rB'$ by one unit but conserves  
the combination $\widetilde{\rB} = \rB + \rB'$. This phenomenon     
can be tested 
via the neutron disappearance ($n\to n'$) and regeneration 
($n \to n' \to n$)  experiments \cite{BB-nn',More,Pokot,HFIR}.
 
As it was shown in Ref. \cite{BB-nn'},  the phenomenological and cosmological bounds
do not exclude $nn'$ mixing mass to be as large as $\eps \sim 10^{-15}$~eV, 
which corresponds to few seconds for the characteristic oscillation time ~$\tau = 1/\eps$.\footnote{In  
this paper we use natural units $c=1$ and $\hbar=1$.}   
In fact,  for free neutrons $n-n'$ oscillation  is suppressed by 
the medium effects as the presence of matter and magnetic fields  \cite{BB-nn',More}.      
But in deep cosmos it could proceed without suppression,  
with interesting implications e.g. for the  cosmic rays at super-GZK energies, 
as far as the neutrons produced in $p\gamma \to n \pi^+$ reactions on relic photons 
can promptly oscillate into $n'$ and then decay in M sector, $n'\to p' e'\ov{\nu}'$  \cite{UHECR,UHECR1}. 
Several dedicated experiments 
searching for $n-n'$ oscillations still allow rather short oscillation times 
\cite{Ban,Serebrov1,Altarev,Bodek,Serebrov2,ILL,Abel}. 
Moreover, some of their results show deviations from null hypothesis \cite{Ban,ILL,Nesti}  
and new experiments are underway for testing these anomalies \cite{Broussard,ESS}. 

 
It is remarkable that the nuclear stability yields no limit on  $\eps$ 
since  $n\to n'$ conversion in nuclei 
is forbidden by the energy conservation \cite{BB-nn'}. 
However,  in the neutron stars (NS)  
$n\to n'$ conversion is energetically favored, and it can gradually  transform 
the NS into mixed stars partially consisting of M matter. 
Since in super-dense nuclear medium  $n-n'$ oscillation is strongly suppressed,   
for $\eps \sim 10^{-15}$~eV the effective conversion time appears to be much 
larger than the universe age \cite{BB-nn'}. 
Nevertheless,  it can have observable effects  for the NS which were
analyzed in details in Ref.~\cite{BBMT}.
Various aspects of $n-n'$ oscillation in the NS were discussed also
in Refs.~\cite{INT,Massimo,LHEP,Nussinov-new,Ciancarella,Makakin}.

By the essence of the mirror matter, the neutron stars should exist also in dark M sector.  
In fact, M baryons represent DM of asymmetric type, 
 with its abundance determined by  the baryon asymmetry (BA) in M world.  
However, a priori the sign of mirror BA is unknown.\footnote{Notice that 
by naming $n'$ as mirror neutron, we implicitly extend the notion of our baryon charge $\rB$ 
to that of M species $\rB'$ classifying the latter via the combined charge 
$\widetilde{\rB} = \rB + \rB'$  which is conserved by $nn'$ mixing \eqref{n-npr}. 
In fact,  the M species with $\widetilde{\rB}=1$ and $\widetilde{\rB}=-1$  for us  respectively are 
the mirror baryons (MB) and anti-mirror baryons (AMB)  
(namely, $\ov{n}'$ is anti-mirror neutron or mirror antineutron), 
no matter how  they are qualified by the inhabitants of M world. }   

The sign of ordinary BA, ${\cal B} = \text{sign}(n_b - n_{\bar{b}})=1$, 
is fixed by some baryogenesis mechanism 
which created the primordial excess of baryons over antibaryons.  
At first glance, the sign of mirror BA  should be the same by M parity, 
${\cal B}' = \text{sign}(n_{b'} - n_{\bar{b}'})=1$.  
This would be the case if the identical baryogenesis mechanisms 
act {\it separately} in the O and M sectors. 
However,  one can also envisage a  co-baryogenesis scenario e.g. 
via the processes which violate $\rB$ and $\rB'$ but conserve $\widetilde{\rB} = \rB + \rB'$
(which processes can in turn be related to underlying new physics  which induces  
$nn'$ mixing \eqref{n-npr}).     
In this case, for  ${\cal B}$ being positive, the null O+M asymmetry in the Universe  
would imply the negative ${\cal B}'$. 
The co-genesis models 
which  induce BA's in both sectors via the $\rB\!-\!\rL$ and $\rB'\!-\!\rL'$ violating 
cross-interactions between the O and M particle species   
indeed predict ${\cal B}$ and ${\cal B}'$ of the opposite signs \cite{ADM,ADM-IJMP}. 

Hence, depending on the sign of ${\cal B}'$, the compact stars in M sector 
can be the mirror neutron stars (MNS) or anti-mirror neutron stars (AMNS)  
consisting respectively of the MB or AMB.   
Correspondingly, $n'\to n$ transitions in the MNS (or $\ov{n}' \to \ov{n}$ in the AMNS) 
can create  the cores of ordinary matter (or antimatter) in their interior.  
In both cases these cores can be detectable as hot compact sources emitting the photons 
in the far UV and soft X-ray ranges \cite{BBMT}, 
but this electromagnetic emission cannot trace their composition (matter or antimatter).  
However,  the presence of  interstellar medium makes the difference.
%
Namely, accretion of interstellar gas on the MNS with O matter core  
can only heat it and induce X-ray emission. However, in the case of AMNS,  
the accreted gas will annihilate on 
the antimatter core producing gamma-rays with energies up to a GeV or so. 

Interestingly, the recent analysis~\cite{Dupourque:2021poh} based on the 10-year 
 Fermi Large Area Telescope (LAT)  gamma-ray source catalog \cite{Fermi-LAT:2019yla}
has identified 14 point-like candidates emitting $\gamma$'s with a spectrum 
compatible to baryon-antibaryon annihilation \cite{Backenstoss:1983gu}, 
and not associated with objects belonging 
to established gamma-ray source classes.  
 
In this letter we discuss the possibility whether the unusual sources of this type can 
be the AMNS with the antimatter cores  which produce the annihilation $\gamma$-rays
by accretion of interstellar gas, and whether some part of this antimatter can escape 
from the AMNS producing the antinuclei detected by AMS-02 in cosmic rays. 



\medskip 

\noindent
{\bf 2.}  
Neutron stars are presumably born after the supernova (SN) explosions of massive 
progenitor stars followed by the neutronization of their iron cores.  
The NS are formed with very high rotation speed 
and with the surface magnetic fields as large as  $10^8\div10^{15}$ Gauss. 
This makes many of them observable as pulsars due to their electromagnetic radiation.  
The NS are believed to have a compact onion-like structure: the inner core of dense nuclear liquid  
dominantly consisting of neutrons (with less than 10\,\% fraction of protons and electrons), 
and the outer shell consisting of heavy nuclei which form a rigid crust at the surface 
(for reviews, see e.g. \cite{Lattimer}).
The  NS mass-radius ($M$--$R$) relations depend  on the equation of state (EoS), 
i.e. the pressure--density relation in dense nuclear matter. 
The masses of the known pulsars range within $M=(1\div 2)\,M_\odot$, 
and the observation of $2 M_\odot$ ones disfavors the too soft EoS. 
E.g. the Sly  EoS \cite{Sly} allows the maximum mass $M_\text{max} = 2.05\,M_\odot$, 
while some of the realistic EoS reviewed in  Refs. \cite{Ozel} can afford somewhat larger masses, up to 
$2.5\,M_\odot$ or so.    All these EoS predict the NS radii  
in the range $R\simeq (10\div 15)$~km.  

The baryon amount $N$ is related to the NS mass as 
\be{M-N} 
 N = \kappa M/m \simeq \left(\frac{\kappa M}{1.5\,M_\odot}\right) \times 2\cdot 10^{57} 
\ee
where $m$ is the nucleon mass, 
and $\kappa \simeq 1$ is the EoS dependent factor which, for a given EoS, 
mildly depends also on $M$.  
E.g. the SLy EoS implies $\kappa\approx 1.1$ for the typical NS masses $M \simeq 1.5\,M_\odot$, 
increasing to $\kappa\approx 1.2$ for heavy NS with $M \approx 2\,M_\odot$ \cite{Sly}. 
The deficit between the gravitational mass $M$ and equivalent baryonic mass $M_b =m N_b$,  
$M_b -M = (\kappa-1) M $, corresponds to the gravitational binding energy.

In principle, 
at extremal densities the energetically favored ground state
can be strange quark matter \cite{Bodmer}. 
Therefore, the NS which mass can reach a critical value by accretion from the companion 
or by the matter fall-back after the SN explosion can be transformed \cite{BBDFL} 
into hybrid stars (HS) with more onion-like structure including  the quark matter core, 
or even entirely quark stars (QS), depending on the amount of accreted mass and  
on the details of the quark matter EoS. 
Interestingly, the bimodal distribution of the pulsar masses \cite{Schwab}
can indeed correspond two different types of stars, 
the lighter stars being the NS and heavier stars being the HS/QS.  
In the following, we shall concentrate on the NS, 
addressing the possibility of the HS/QS only in proper occasions.

 \medskip 
 \noindent {\bf 3.}
The  core-collapse of O star should produce a NS  entirely 
consisting  of  ordinary nuclear matter.\footnote{Modulo some tiny amount of M matter 
which can be accreted by the progenitor star during its lifetime, or produced 
 via $n-n'$ oscillation of the neutrons involved in certain chains of 
nuclear reactions at late stages of its evolution before the core-collapse.} 
But once the NS is formed, then $n\to n'$ transitions in its liquid nuclear core 
can effectively produce M neutrons.\footnote{The NS liquid core is dominated by the neutrons 
and we shall neglect subdominant fraction of protons for simplicity. 
In the crust the neutrons are bound in heavy nuclei and  for them $n-n'$ conversions 
is ineffective, but this fraction is also negligible. 
}

 
The oscillation $n-n'$ in nuclear medium 
is described by the Schr\"odinger equation with effective Hamiltonian
\be{H-osc}
H = \mat{ \Delta E }{\eps} {\eps} { 0 } 
\ee
where the off-diagonal term $\eps$ comes from $n-n'$ mass mixing \eqref{n-npr}, 
and $\Delta E$ is the medium-induced energy splitting  between  $n$ and $n'$ states.
It can be taken as the difference  of the in-medium optical potentials induced by 
their coherent scatterings, $\Delta E = V-V'$. 
Namely, for O neutrons we have $V=2\pi a n_b/m$, where $m$ is the neutron mass, 
$a= a_3 \times 3$~fm  is its scattering length, and $n_b = \xi \times 0.16~{\rm fm}^{-3}$ 
is the baryon density,  i.e. $\xi$ is $n_b$ in units of nuclear density.
If the M baryon density in the star is small, $n'_b \ll n_b$, $V'$ can be neglected    
and we get $\Delta E = V =  a_3  \xi  \times 125\,\text{MeV}$.\footnote{Zeeman energy 
induced by the neutron magnetic moment $-\mu = 6\times 10^{-12}$\,eV/G  
is also negligible since $\vert \mu B \vert < 10^{-2}$~MeV 
 even in magnetars where the magnetic field $B$ can reach $10^{15}$\,G or so. }
 Hence,   in a dense nuclear medium $n-n'$ mixing angle is 
$\theta = \eps/V \simeq (\eps_{15}/ a_3 \xi) \times 8\cdot 10^{-24}$ 
 ($\eps_{15} = \eps \times 10^{15}\,{\rm eV}^{-1}$), 
 and the mean probability of 
 $n-n'$ oscillation $P_{nn'} = 2\theta^2$ is extremely small.  
 
 The mirror neutrons are mainly produced via the scattering process $nn \to n'n$.\footnote{Also 
 $np \to n'p$ scatterings work, but we neglected them  since the proton fraction 
 at the typical densities of the liquid core is just of few per cents.}
  Its rate  can be estimated as $\Gamma = 2\theta^2 \eta \langle \sigma v \rangle  n_b$,  
 where $\sigma = 4\pi a^2$ is the $nn$ elastic scattering cross-section and $v=p_F/m$ is the mean relative 
 velocity, while the  factor $\eta \approx 0.18$ takes into account the Pauli blocking implying 
that in $nn \to nn'$ process the final state $n$ should have the momentum above the Fermi 
 momentum $p_F = \xi^{1/3} \times 340$~MeV.  
Hence, dependence on the scattering length cancels out and we obtain  \cite{BBMT}: 
 \be{Gamma-0} 
\Gamma  = \eps_{15}^2 \,\xi^{-2/3}  
\times 3 \cdot 10^{-47} ~ \text{GeV}
 \ee 
This rate depends on the baryon density $n_b$ which varies by an order of magnitude 
from the NS centre to the outer edge of its liquid core, so that  
in dense central regions  $n-n'$  transition is somewhat slower
than in peripheral regions.  
Hence,  the factor $\xi$ in \eqref{Gamma-0} should be 
averaged over the baryon density profile in the star. Taking $\langle n_b \rangle = N/V$ 
where $V$ is the NS volume and $N=\kappa M/m$ is  its total baryon number, 
for the effective time $\tau_\eps = \langle \Gamma\rangle^{-1}$ of $n-n'$ conversion    
we get: 
\be{taueps} 
 \tau_\eps = \frac{a_{R}}{\eps_{15}^{2}} \,\left(\frac{M}{1.5\,M_\odot}\right)^{2/3}  \times 10^{15} ~{\rm yr}
\ee 
where factor $a_R = \kappa^{2/3}/R_{12}^2\simeq 1$, with $R_{12}= R/12\,{\rm km}$   
normalized to the typical NS radii,   
mildly depends on the EoS which determines $M$--$R$ relation for the NS. 
As we see,  $\tau_\eps$ becomes comparable  
to the universe age $t_U=14$~Gyr for $\eps = 4\times 10^{-12}$~eV or so. 

Thus, by integrating the Boltzmann equations 
\be{Boltzmann}
\dot{N}'=  \langle \Gamma \rangle N ,  
\quad\quad   \dot{N} = -\langle \Gamma \rangle N  
\ee
and observing that the overall O+M baryon number $N+N'$ is conserved, 
for the fraction of M baryons 
produced in the NS of the age $t<t_U$ we get:  
 \be{Nprime}
\frac{N'}{N+N'} = \frac{t}{\tau_\eps} 
= 10^{-5} \times \frac{\eps_{15}^2}{a_{R}} \left(\frac{1.5\,M_\odot}{M}\right)^{\!2/3} \!\!\left(\frac{t}{10\,{\rm Gyr}}\right) 
\ee
Thus, for $\eps \sim 10^{-15}$\,eV the amount of M baryons in the oldest NS 
are comparable to their amount in planets. 

Due to the Pauli blocking, the process $nn \to nn'$ process takes place 
for the neutron momenta close to the Fermi surface, so that 
$n'$ is typically produced with the Fermi energy
$E_F = p_F^2/2m = \xi^{2/3} \times 60$~MeV.  
Hence, the energy production rate per baryon  is independent of $\xi$: 
$\Gamma E_F \simeq \eps_{15}^2  \times 3 \cdot 10^{-24}$~{\rm GeV/s}. 
Then, multiplying this by the baryon number in the NS  \eqref{M-N}, 
 for the total rate of the energy production we obtain
\be{E-dot} 
\dot{\cal E} \simeq \Gamma E_F N \simeq \eps_{15}^2 \left(\frac{M}{1.5\,M_\odot}\right) 
\times  10^{31}~\text{erg/s} 
\ee 
which energy will be radiated away via the mirror photons and neutrinos.   

In the case of HS/QS situation is somewhat different. 
Transition $n-n'$   should be ineffective in quark matter  since there are not much 
neutrons. In addition, quark matter is thought to be self-bound \cite{Bodmer}, 
in particular if it is in color superconducting phase \cite{Alford}, 
and its transition to M matter is not energetically favored. 
Hence in the QS, almost entirely consisting of quark matter, 
$n-n'$ transitions play practically no role. In the hybrid stars  
$n-n'$ transition should take place in its shell consisting of the neutron liquid. 
Thus, in principle  the HS can also develop the M matter cores in their interior. 
However, the energy production rate  in the HS,  
with less neutrons employed, 
can be much lower than it was estimated in Eq. \eqref{E-dot}  for the NS. 
 
For the evolution time $\tau_\eps$ being much larger than the cooling time,  
Eq.~\eqref{E-dot} in fact determines the NS mass loss rate due to $n-n'$ conversion:
\be{M-dot} 
\dot{M}/M 
\simeq -\eps_{15}^2 \times 10^{-16}~{\rm yr}^{-1}  
\ee
The mass loss changes the orbiting period in the compact NS binaries 
as $\dot{P}_b/P_b \simeq -2\dot{M}/M$ \cite{Nussinov-new}.  
 The observational data on the orbital period decay in some well-studied compact 
binaries (as e.g. the famous Hulse-Taylor pulsar B1913+16) yield  an upper bound 
roughly  as $\eps < 10^{-13}$~eV \cite{Nussinov-new,BBMT}.  
Somewhat stronger bounds,  $\eps< 10^{-15}$~eV or so \cite{BBMT}, 
can be obtained from the surface temperature measurements  of old pulsars, 
which for PSR J2144$-$3933, with $T_\text{surf} < 4\times 10^4$ K \cite{Guillot}, 
approaches $\eps < 10^{-16}$~eV. But this object, being an isolated slow pulsar 
with characteristic age $\tau_c\simeq 0.3$~Gyr and {\it unknown} mass, could 
in principle be quark matter dominated.  In this case the above limit would be invalid. 
More generally,  the observational determination of the NS surface temperature 
is influenced by environmental factors as composition of the pulsar magnetosphere and 
interstellar extinction, and its theoretical interpretation depends on the cooling models. 
Therefore, in the following we conservatively stick to benchmark values $\eps \sim 10^{-15}$\,eV. 
 
 M baryons produced in the NS will form  a mirror core 
in its interior. 
In fact, $n'$ produced at the first instances after the NS formation will undergo $\beta$-decay 
$n' \to p'e'\bar\nu'$ forming a plasma consisting of $p'$ and $e'$ components. 
But very soon, in hundred years or so, 
the density produced M matter $n'_b$ will exceed $10^{26}/{\rm cm}^{3}$, 
equivalent to the baryon density in the sun's centre.  
In gravitationally captured hot M matter nuclear reactions will be ignited
among $p'$ and $n'$ continuously produced in the NS,  
forming the M nuclei.  In the very old NS where the central density of M baryons
could reach $n_{b'} = 0.07$~fm$^{-3}$ or so, they can undergo a "neutronization" 
and form a degenerate liquid core dominantly consisting of $n'$. 

The energy production rate \eqref{E-dot} can be substantial: e.g. for $\eps_{15}=10$ it would 
comparable to the solar luminosity ${\cal L}_\odot = 4 \times 10^{33}$ erg/s. 
It should be radiated away by M photons and M neutrinos,   
but given the complexity of nuclear reactions in hot core, 
it is difficult to precisely determine the energy fractions emitted in two channels.  
In any case, the fraction $x$ emitted in thermal M photons should be significant,  
and the ``mirror photosphere" temperature in the NS   can be estimated  
by equating $x \dot{\cal E} = 4\pi R_\text{ph}^2 \times \sigma T^4$,
where $R_\text{ph}$ is the photosphere radius and $\sigma$ is the Stefan-Boltzmann constant:
\be{surface-T}
T = \eps_{15}^{1/2} \left(\frac{x \kappa M}{1.5\,M_\odot}\right)^{1/4} \left(\frac{1~{\rm km}}{R_\text{ph}}\right)^{1/2}   
\times 10^{6} ~\text{K} 
\ee
For determining the observable temperature $T_\infty$, 
the surface redshift effect should be taken into account.

\medskip 
\noindent {\bf 4.} 
Let us reverse now the situation and consider neutron stars in M sector. 
As we have anticipated in the introduction,  
 the BA ${\cal B}'$ in M world can be positive or negative, 
 depending on the baryogenesis mechanism. 
 Hence, if  ${\cal B}'>0$ all compact mirror stars should be the MNS, 
 whereas if  ${\cal B}' < 0$  they all should be the AMNS.  
As far as the O and M sectors have the same microphysics by mirror parity, 
the EoS describing the O and M nuclear matters should be identical, 
and, needless to say,  it should be 
identical for the nuclear and antinuclear matters by C invariance of strong interactions. 
In other words, the NS, MNS and AMNS should have the same $M$--$R$ relations. 

In the absence on $n-n'$ mixing 
these will be dark compact objects, sort of solar mass MACHOs which can be 
detected by microlensing, but  ordinary observer cannot distinguish between the
MNS and AMNS.     If $n-n'$ mixing is switched on, 
then $n'\to n$ transitions will create O matter in the MNS interior, 
while in the AMNS $\ov{n}' \to \ov{n}$ will take place forming the O antimatter.     
These transitions with the effective time and energy production rate  given again
by Eqs.~\eqref{taueps} and \eqref{E-dot},  would form hot cores 
which can be visible for us as bright compact stars with high temperatures \eqref{surface-T}. 
Clearly, this will not allow to determine whether the core is composed of O matter or O antimatter.  
However, the two cases can be discriminated by the accretion of ordinary gas
which, in the case of the AMNS, will annihilate with the antibaryons in its interior.  
Thus, detection $\gamma$-ray sources with a typical baryon-antibaryon annihilation spectrum 
\cite{Backenstoss:1983gu} can be the way to determine the BA sign in mirror world.  


Let us consider an AMNS  of a typical mass $M$ and radius $R$.  
Transitions $\ov{n}' \to \ov{n}$ will produce O antimatter in its liquid core. 
The production rate (antibaryon per second) can be estimated as  
\be{Nbar-dot} 
\dot{N}_{\ov{b}}  = \frac{N_{\ov{b}'}}{\tau_\eps} \simeq \eps_{15}^2 \left(\frac{M}{M_\odot} \right)^{2/3} 
\times 3 \cdot 10^{34}~ \text{s}^{-1}
\ee 
where the amount of AMB in the AMNS $N_{\ov{b}'}$ is given by Eq. \eqref{M-N},  
and  the conversion time $\tau_\eps$ is given by Eq. \eqref{taueps}. 
For simplicity, we put $\kappa =1.1$ and $R=12$~km. 
 
On the other hand, the AMNS will accrete O gas  while it travels in 
diffuse interstellar medium (ISM).  The accretion rate (baryon per second) reads 
\be{N-dot}
\dot{N}_b  \simeq \frac{(2G M)^2  n_\text{is}}{v^3}  
\simeq \frac{10^{32} }{v_{100}^{3} } \times   
\left(\frac{n_\text{is}}{1/\text{cm}^{3}} \right) \left(\frac{M}{M_\odot} \right)^2\! \text{s}^{-1}
\ee
where $n_\text{is}$ is the baryon density in the ISM 
and $v= v_{100}\times 100$~km/s is the  star velocity relative to the ISM 
which is normalized taking into account that typical kick velocities of pulsars are order 
100 km/s.  
 
The ratio between the two rates 
\be{ratio} 
\frac{\dot{N}_{\ov{b}}}{\dot{N}_b} \simeq \eps_{15}^2 v_{100}^3 
\left(\frac{1/\text{cm}^{3}}{n_\text{is}} \right) \left(\frac{M}{M_\odot} \right)^{-4/3} \times 300
\ee
depends, for a given $\eps$,  on the AMNS mass $M$ 
(which can range within $M=(1\div2)\,M_\odot$ as that of the normal NS) 
and, more critically, on its velocity $v$ and on the ISM density. 
The velocity distribution of observed pulsars seems to be bi-modal: a part of 
pulsars have kick velocities $v > 100$ km/s and some achieving 1000 km/s, 
while  others probably receive a very little kick -- otherwise they would not be contained 
in globular clusters which have small escape velocities. 
The origin of pulsar kicks is largely unknown, but we consider that this bimodal 
distribution applies also to the AMNS or MNS.  

For definiteness, let us take $\eps\sim 10^{-15}$ eV as a benchmark value. 
For the AMNS with $v > 100$~km/s we have $\dot{N}_{\ov{b}}/\dot{N}_b \gg 1$,   
i.e. the antibaryon production rate \eqref{Nbar-dot} 
is much larger than the baryon accretion rate \eqref{N-dot}. 
In this case  the antimatter core can be formed and it will emit O photons 
 with the energy luminosity \eqref{E-dot},  
 Such cores can be observed as bright point-like sources in the far UV and soft X-ray ranges. 
 As for the accreted baryons, they should annihilate on the core surface. 
 The annihilation photons will be produced with the rate 
  $L_\gamma = l_\gamma \dot{N}_b$, where $l_\gamma \approx 4$ 
 is the average multiplicity of $\gamma$'s per $p\ov{p}$ annihilation \cite{Backenstoss:1983gu}. 
 The rate of energy production is $2 m \dot{N}_{b}$,    
 about a half of which contributes in heating the core 
 (in addition to \eqref{E-dot}), and another half will be emitted 
 from its surface in $\gamma$-rays. Hence, 
 the energy flux from such a source at  a distance $d$ 
 will be $J\simeq m \dot{N}_{b}/4\pi d^2$, or numerically 
   \be{J-ann}
J  \simeq \frac{10^{-12} }{v_{100}^{3} }    
\left(\frac{n_\text{is}}{1/\text{cm}^{3}} \right) \! \left(\frac{M}{1.5\,M_\odot} \right)^2 \!
\left(\frac{50\,\text{pc}}{d} \right)^2 \! \frac{\rm erg}{{\rm cm}^2 {\rm s} } 
\ee
For the AMNS  travelling with $v> 100$ km/s  in the Milky Way (MW)  
this emission can be below the diffuse $\gamma$-background and the source 
may remain unresolved unless this source is closer than 50 pc or so. 
However, if the AMNS has less velocity, say $v\simeq 30$ km/s, 
and it incidentally crosses a high density region 
as e.g. cold molecular cloud with $n_\text{is} > 10^3/\text{cm}^{3}$, 
the observability distance can be increased up to several kpc. 

On the other hand, for the slow AMNS moving in galactic discs with  $v < 10$ km/s, 
the antibaryons produced in its interior can be outnumbered by the accreted baryons, 
i.e. $\dot{N}_{\ov{b}} < \dot{N}_b$. In this case the antimatter core canot be formed and 
the thermal emission with temperature \eqref{surface-T} will be suppressed. 
The produced antibaryons will readily annihilate with the already accreted baryons 
and the $\gamma$-ray luminosity will be $L_\gamma= l_\gamma \dot{N}_{\ov b}$. 
In other words, the $\gamma$-luminosity of the object 
will be defined by the lesser value between $\dot{N}_{\ov{b}}$ and  $\dot{N}_b$. 
Namely, if the ratio $\dot{N}_{\ov{b}}/\dot{N}_b$ \eqref{ratio} is less than one, 
then instead of Eq.~ \eqref{J-ann} we would have for the energy flux 
   \be{J}
J  \simeq  10^{-12}  \times  \left(\frac{\eps_{15}}{0.05} \right)^2 \!  
  \left(\frac{M}{1.5\,M_\odot} \right)^{2/3}\!
\left(\frac{50\,\text{pc}}{d} \right)^2 \! \frac{\rm erg}{{\rm cm}^2 {\rm s} } 
\ee
where the $\eps$-dependent factor is normalized to indicate that  for 
$\eps < 5 \times 10^{-17}$~eV such a sources become too faint 
to be resolved at distances larger than 50 pc or so.  

The search of $\gamma$ -ray sources with a spectrum compatible with baryon-antibaryon annihilation 
 was recently performed in Ref.~\cite{Dupourque:2021poh}. 
Analyzing 5787 sources  included in 4FGL catalog \cite{Fermi-LAT:2019yla}
based on 10 years of observations with the Fermi LAT, 14 candidates were found  
which were selected by applying the following criteria: \\
(i) extended candidates were excluded (with angular size larger than the LAT resolution 
at energies $E < 1$~GeV); \\
(ii) sources associated with objects known from other wavelengths and belonging to 
established $\gamma$-ray sources were excluded, as e.g. pulsars and active galactic nuclei; \\
(iii) sources with significant higher energy tail above a GeV were excluded since 
the  baryon-antibaryon annihilation $\gamma$-spectrum should end up at the nucleon mass.  

Interestingly, the distribution of the sources in the sky shown 
in Fig. 1 of Ref.~\cite{Dupourque:2021poh}
 very much resembles the distribution of observed pulsars. 
Only two candidates have galactic coordinates compatible with the MW disc, 
while the 11 candidates having galactic latitudes  $\vert b \vert > 10^\circ$ 
(among which 7 candidates with $\vert b \vert > 30^\circ$) can be assigned to the MW halo.
Interestingly, the sources belonging to the disc are the brightest, with the energy 
fluxes $J \geq 10^{-11}$ erg cm$^{-2}$ s$^{-1}$, while the ones with higher galactic 
latitudes become increasingly fading, and the source J2330-2445 ($b=-71,7^\circ$)
is the taintest, with $J < 2\times 10^{-12}$ erg cm$^{-2}$\,s$^{-1}$.  
In view of Eq. \eqref{N-dot}, this may well explained by correlation of the accretion rate 
with the distribution of the ISM densities, though the velocity distribution of the stars 
remains the key issue. 

Probably it is too early to claim the discovery. These sources are  
all faint, close to the Fermi LAT detectability threshold, and they may well belong 
to a known $\gamma$-ray source classes, or maybe mimicked  by imperfections 
of the background interstellar emission. In fact, the authors of Ref.~\cite{Dupourque:2021poh}
take a conservative attitude and translate their findings into an upper limit 
on the local fraction of such objects with respect to normal stars. 
The clear  identification of these sources  is very challenging, 
and requires serious multiwavelength search. 

The possibility of mirror NS (or HS) being the engines of our antimatter can have interesting 
implications since they can produce antinuclei in the ISM. 
Namely, for $\eps \sim 10^{-15}$~eV,  transitions $\ov{n}'\to \ov{n}$ transition in the AMNS 
produces about $10^{52}$ antibaryons forming the hot and dense core in which 
nuclear reactions  should form the antinuclei. 
However, these antinuclei  are gravitationally trapped and 
the question is how they can escape from the star. This possibility can be provided by the 
mirror neutron star mergers.  In the coalescence of the two AMNS, their small antimatter cores 
do not merge at the same instant but continue the orbiting and then explode due to the decompression 
producing a hot cloud of the neutron rich antinuclei.  Most of these antinuclei, being stable 
only at the extreme densities, will decay into the lighter ones which are stable in normal 
conditions.  Therefore, the antinuclei produced by such ``sling" effect can leave the coalescence site 
and propagate in the outer space. In addition, they can have some additional acceleration 
if reasonably large magnetic fields are formed the rotating ordinary anti-core 
during its evolution before the merger.   
The AMS-02 experiment hunting for the antinuclei in the  cosmic rays has reported, 
as the preliminary results, the detection of  eight antihelium events  (among which 
two are  compatible with antihelium-4 and the rest with  antihelium-3), 
The  fraction  $\sim 10^{-8}$ of antihelium with respect to measured fluxes of the helium 
 is too high to be explained by the conventional production mechanisms. 
Interestingly, the rate of the NS mergers $\sim 10^3$~Gpc$^{-3}$ yr$^{-1}$, with  
$\sim 10^{52}$ antibaryons produced per a merger,  is nicely compatible with 
this fraction of antihelium.  In addition, our mechanism should produce  also the heavier antinuclei,  
and thus AMS-02 can be the place where to find fantastic animals as anticarbon or antioxygen   
which would be a key for many mysteries.

 \medskip
\noindent {\bf 5.} 
 %
Let us discuss our scenario in wider context, and have a pleasant walk and pleasant talk 
with the Walrus and the Carpenter, 
viewing and reviewing panorama of two parallel worlds.
This picture is based on the direct product $G\times G'$ 
of two identical gauge groups represented by the SM (or some its extension), 
with the total Lagrangian  
 \be{Lagr}
{\cal L}_{\rm tot} = {\cal L} + {\cal L}' + {\cal L}_{\rm mix}  
\ee
 where ${\cal L}$ describes the O matter and  ${\cal L}'$ the dark M matter,  
 while the mixed Lagrangian ${\cal L}_{\rm mix}$ 
 describes the possible cross-interactions between the particles of two sectors. 
 The identical forms of ${\cal L}$ and ${\cal L}'$ is ensured by  
a discrete symmetry $G\leftrightarrow G'$ interchanging  all O species: 
the fermion, gauge and Higgs fields of $G$ sector,  with their M partners:  
the fermion, gauge and Higgs fields of $G'$ sector. 
In the context of extra dimensions, 
it can be viewed as a geometric symmetry between 
two parallel 3-branes on which the O and M particle species are localized. 

M baryons as cosmological DM 
have specific cosmological implications  \cite{Khlopov,BDM,BCV,IV,BCCV,BCCP}. 
%
Although O and M components  have identical microphysics,  
their cosmological realizations cannot be quite different. 
Namely, the viability of M sector 
requires the following conditions \cite{BDM,BCV}: \\
--  {\it Asymmetric reheating}: after inflation the O and M sectors are  reheated asymmetrically, with $T>T'$, 
which can naturally occur in certain models; \\
-- {\it Out-of-equilibrium:}  interactions between O and M particles are  feeble enough 
in order to maintain the initial temperature asymmetry in  subsequent epochs. 
In other words, all cross-interactions in ${\cal L}_\text{mix}$ 
should remain out-of-equilibrium at any stage after inflation, at least before the 
Big Bang Nucleosynthesis (BBN); \\
-- {\it No extra heating:}  after inflation both sectors evolve almost adiabatically 
and the temperature difference $T' < T$ is not erased  by entropy production in M sector
due to possible 1st order phase transitions.  
 
  Namely, the BBN bounds demand $T'/T < 0.5$ \cite{BCV} while the post-recombination cosmology 
 is yet more restrictive, 
requiring $T'/T < 0.2$ or so \cite{BCCV}. This has interesting consequences  
for the primordial chemical content in M world: while O world is dominated by hydrogen, 
with it primordial mass fraction being 75\,\% and that of helium being 25\,\%, 
M world should be helium dominated, with the mass fractions of M hydrogen and M helium 
being respectively  25\,\% and 75\,\% or so \cite{BCV}. 

Along with the ordinary stars also dark M stars can be formed the Galaxy. 
However, since mirror world is colder, 
the first M stars should start to form somewhat earlier than the first  (population III) stars in O sector.  
M stars, being helium dominated, should have somewhat different initial mass function, 
and their evolution should be more rapid as compared to O stars \cite{BCCP}. 
Since helium is less opaque than hydrogen,  
the massive M stars should suffer less mass losses due to stellar winds 
end up their life collapsing directly into black holes (BH).  
The intermediate mass M stars can explode as SN and form mirror neutrons stars,  
while the solar mass M stars as white dwarfs  can survive until present times. 
 All these objects can constitute a dominant fraction of dark matter in the Galaxy. 
 Namely, the galactic halo can be viewed as elliptical mirror galaxy of M stars and BH  
 in which O matter forms the disc \cite{BCV}.  
 M matter could contribute also to the disc, but the density of M stars in the disc 
 should not exceed the density of O stars  \cite{Roux}. 
 All these objects can be observed via microlensing  as the Machos  in different mass ranges. 
 The present limits from EROS-MACHO observations 
 do not exclude the possibility  of the galactic halo dominated by dark objects as BH with masses 
$M > (10 \div 100)\,M_\odot$ while the fraction of dark stars with $M\sim M_\odot$ can be 
$\sim 10\%$  or so.  
This proportion can correspond to the abundance of the LIGO gravitational wave (GW) signals 
\cite{LIGO}
from the BH mergers with typical masses $M\sim (10\div 50)\,M_\odot$ or so  \cite{Merab}. 
In addition, some of the peculiar LIGO events with one or both light components and 
no optical counterpart can be viewed as the BH-MNS or MNS-MNS mergers  \cite{Merab1}.    

Now the time has come to talk of many things: of shoes and ships and sealing-wax, 
of cabbages and kings.... 
 there is a subtlety related to the chiral character of the fermion representations in the SM: 
in our weak interactions fermions are left-handed (LH)   
while the antifermions  are right-handed (RH), 
and two systems would be symmetric if not CP-violating effects. 
The value of the BA in the Universe, and in particular its sign ${\cal B} = \text{sign}(n_b - n_{\bar{b}})=1$, 
is fixed by (unknown)  baryogenesis mechanism {\it a l\'a} Sakharov~\cite{Sakh}
that created primordial excess of baryons over antibaryons  
due to CP-violation in some out-of-equilibrium processes 
 violating $\rB$  (or $\rB-\rL$~\cite{KRS}). 
In parallel sector the situation is same, apart of an ambiguity 
in the CP-violation pattern  distinguishing between 
the M fermions and M antifermions, and determiing the sign of mirror BA 
${\cal B}' = \text{sign}(n_{b} - n_{\bar{b}'})$. 
This ambiguity is related to the fact that 
$G\leftrightarrow G'$  symmetry can be realized in two ways: 
{\it with} or {\it without} chirality change between the O and M species~\cite{Alice}. 

Namely, for three families of O fermions $f_{L,R}$ described by $SU(3) \times SU(2) \times U(1)$, 
the left-handed (LH) quarks $q_L=(u_L,d_L)$ and leptons $\ell_L=(\nu_L,e_L)$  
are weak doublets while the right-handed (RH) ones $u_R,d_R$ and $e_R$ 
are singlets. 
The antifermion fields are obtained by complex-conjugation, 
$\bar{f}_{R,L} = C \gamma_0 f_{L,R}^\ast $, 
and they have opposite chiralities:  
$\bar{q}_R=(\bar{u}_R,\bar{d}_R)$ and $\bar{u}_L,\bar{d}_L$ 
are antiquarks, and  $\bar{\ell}_R=(\bar{\nu}_R,\bar{e}_R) $ and $\bar{e}_L$ 
are antileptons. 
The invariance under the transformation CP: $f_{L,R} \to \bar{f}_{R,L}$, 
i.e. symmetry between the fermions and antifermions,
is violated by the complex Yukawa couplings  with the Higgs doublet $\phi$. 
%

As for three analogous families  of  $SU(3)' \times SU(2)' \times U(1)'$ in M sector, 
we invert definitions and denote the species $\ov{f}'_{L,R}$ with the LH weak interactions 
as M antiquarks: 
$\bar{q}^{\prime}_L =(\bar{u}^{\prime}_L,\bar{d}^{\prime}_L)$ 
and $\bar{u}^{\prime}_R,\bar{d}^{\prime}_R$, and M antileptons: 
$\bar{\ell}^{\prime}_L = (\bar{\nu}^{\prime}_L,\bar{e}^{\prime }_L)$ and $\bar{e}^{\prime}_R$. 
Correspondingly, the respective anti-species $f'_{R,L}$ with the RH weak interactions 
we call M quarks: 
$q'_R=(u'_R,d'_R)$ and $u'_L,d'_L$, and M leptons: $\ell'_R=(\nu'_R,e'_R)$ and $e'_L$.   
This is just a convention: we could name them in the opposite way. 
The M fermions and antifermions are equivalent modulo CP violating phases 
in the Yukawa couplings of the mirror Higgs doublet $\phi'$. 

Hence, one can impose a symmetry ${\cal Z}_2:$ $ f_{L,R} \leftrightarrow \ov{f}'_{L,R}$ 
interchanging the twin species of the {\it same chirality}, 
i.e. each O fermion with the corresponding M antifermion. 
Alternatively, we can employ ${\cal Z}^{LR}_2 ={\cal Z}_2 \times \text CP$ 
under exchange $ f_{L,R} \leftrightarrow f'_{R,L}$ 
between the O and M fermions which, in our definition, have the {\it opposite chiralities}.
 (Clearly, both of these transformations should be also 
complemented by a proper exchange between the gauge and Higgs fields of two sectors.) 
 In the former case the M antimatter should have exactly the same CP-violating physics  
as our matter. In the latter case the equivalence holds between the `left-handed' O matter  
and the `right-handed' M matter    
which means that P parity, maximally violated in weak interactions of each sector,
is in some sense  restored between two sectors. 
In fact,  this was the original motivation for introducing mirror fermions~\cite{Mirror,FLV} 
(for a historical overview see Ref.~\cite{Okun}). 
But the real difference is related to CP-violation which was not yet 
discovered at the time of original works~\cite{Mirror}. In the absence of 
CP-violating phases ${\cal Z}_2$ and ${\cal Z}_2^{LR}$ would be equivalent. 

Both of these discrete symmetries ensure the identical form of the O and M Lagrangians 
$\cL$ and $\cL'$ in \eqref{Lagr}, modulo the issues of CP-violation.  
Nevertheless, in the absence of cross-interaction terms $\cL_\text{mix}$ in \eqref{Lagr}, 
with the O and M particles interacting only gravitationally, 
no experiment can discriminate between the two possibilities. 

 On the other hand, there are no fundamental reasons for neglecting the 
cross-interactions in $\cL_\text{mix}$ which in fact are the portals for the mirror DM detection 
and identification. 
For example, the simplest possibility is a kinetic mixing term between the O and M photons,  
$\varepsilon F^{\mu\nu} F'_{\mu\nu}$ \cite{Holdom}. 
The particles of two sectors can interact also through the gauge 
bosons of e.g. the common family symmetry $SU(3)_H$ \cite{PLB} or  
common $U(1)_{\rB-\rL}$ symmetry \cite{ABK}. 
They can also share the Peccei-Quinn symmetry 
and their cross-interaction can be mediated by the axion \cite{BGG}. 
The M gas (dominated by helium, and probably containing some M nuclei of  
carbon-neutrogen-oxygen) can be subject of direct detection via the kinetic mixing 
of the O--M photons or via other portals in $\cL_\text{mix}$ \cite{Cerulli}.

The most interesting interactions in $\cL_\text{mix}$ are the ones 
that violate baryon and lepton numbers of both sectors. 
  
In fact, the conservation of $\rB$ and $\rL$ in the SM  is related to accidental global symmetries 
of the Lagrangian at the level of the renormalizable terms. However, these global symmetries can 
 be broken by higher order operators cutoff emerging from new physics at some large energy scales. 
Namely, $\rL$ should be violated if the neutrino have the Majorana masses, 
 while $\rB$ (or $\rB-\rL$) violation is needed for the baryogenesis. 

In particular, the lowest dimension effective operators which violate the lepton numbers are of 
dimension D=5: 
\be{Ops} 
\frac{A}{M} \big(\ell_L\phi\big)^2 + \frac{A'}{M} \big(\bar{\ell}'_L \phi'\big)^2  
+ \frac{\cal A}{M} \big(\bar{\ell}'_L \phi'\big)\big(\ell_L \phi\big) ~+~ \text{h.c.} %
\ee
where $M$ is the relevant mass scale, and 
$A\!=\!A^T$, $A'\!=\!A^{\prime T}$ and ${\cal A}$ are the matrices of generically complex  
Wilson coefficients  in family space (the family indices as well as $C$-matrix are suppressed). 
The first term, violating $\rL$ by two units ($\Delta \rL=2$), 
after substituting the Higgs VEV $\langle \phi \rangle = v\sim 10^2$~GeV, 
induces the small Majorana masses to the neutrinos, $m_\nu \sim v^2/M$ \cite{Weinberg}.  
The second term ($\Delta \rL'=2$) works analogously in M sector and induces the Majorana 
masses of M neutrinos.   As for the third term, it violates both $\rL$ and $\rL'$ by one unit 
conserving the combination $\widetilde{\rL} = \rL+\rL'$,  and induces the O--M neutrino 
(active-sterile) mixing  \cite{ABS}. 

The first operator in \eqref{Ops} can be induced by seesaw mechanism, by introducing 
the heavy Majorana fermions  $N$ in weak singlet and triplet representations 
which are coupled to the leptons via the Yukawa terms   $Y \ell_L N \phi  + \text{h.c.}$,  
with $Y$ being the matrix of respective coupling constants. 
Then the second operator is induced via the Yukawa terms 
$Y' \bar{\ell}'_L {N}' \phi' + \text{h.c.}$ with the analogous heavy fermions $N'$ of M sector. 
However, there can exist also singlet heavy fermions ${\cal N}$ which are coupled with both 
O and M leptons:  ${\cal Y} \ell_L {\cal N} \phi  + {\cal Y}' \bar{\ell}'_L {\cal N} \phi' + \text{h.c.}$, 
and thus are messengers between two sectors. In this way, 
all three operators in \eqref{Ops} are induced  by the seesaw mechanism, 
and for their coefficients  we get: 
\begin{eqnarray} \label {AApr} 
&& A= Y Z Y^T + {\cal Y} {\cal Z} {\cal Y}^T,   
~~ A'= Y' Z' Y^{\prime T} + {\cal Y}' {\cal Z} {\cal Y}^{\prime T}, \nonumber \\  
&& \qquad \qquad \qquad \widetilde{A} = {\cal Y'} {\cal Z} {\cal Y}^T 
\eeqn 
where  the coefficient matrices $Z$ etc. parametrize the 
inverse mass matrices of $N$, $N'$ and ${\cal N}$ fermions respectively as $Z/M$, $Z'/M$ 
and ${\cal Z}/M$. 
Without losing generality, these matrices can be taken to be diagonal and real, 
while the symmetry under $N \leftrightarrow N'$  implies $Z'=Z$. 

Now the question comes to the sign of BA in M sector. 
Clearly, this depends on the baryogenesis models, for which 
two possible realizations of the discrete inter-sector symmetry, ${\cal Z}_2$ and  ${\cal Z}_2^{LR}$, 
have different implications. 

Let us consider the case  of a baryogenesis mechanisms acting separately in O and M sectors 
in identical manner.  
For example, the BA's can be induced by means of electroweak baryogenesis (EWB) 
in both sectors \cite{BCV}.\footnote{Certainly,   in the SM  context the EWB cannot work
but one can consider e.g. two Higgs doublet or supersymmetric extension.} 
Then, in the case of ${\cal Z}_2$, which yields the CP-violation pattern 
for the M antimatter identical to that of the O matter, the BA's ${\cal B}$ and ${\cal B}'$ 
induced in two sectors should have the opposite signs.   
Alternatively, we can consider a leptogenesis scenario related to the seesaw mechanism,  
due to the CP-violation in the decays of the  heavy Majorana fermions  
$N\to \ell_L \phi$ and $N' \to \bar{\ell}'_L \phi'$ (and analogous decays 
of ${\cal N}$-fermions coupled to both sectors).  
Then ${\cal Z}_2$ ($\ell_L \leftrightarrow \ov{\ell}'_L$)  implies $Y'=Y$ and  ${\cal Y}'={\cal Y}$ 
for the Yukawa couplings in \eqref{AApr}, and so for ${\cal B}$ being positive, 
${\cal B}'$ should be negative. 

As for ${\cal Z}^{LR}_2$ ($\ell_L \leftrightarrow \ell'_R$), it implies the equivalent CP-violation 
between the O and M fermions, and the above mechanisms of separate baryogenesis mechanisms 
in this case predict the same signs of ${\cal B}$ and ${\cal B}'$.  

However, the BA of opposite signs between the O and M sectors can be generated also 
in the case of ${\cal Z}^{LR}_2$ symmetry. 
This occurs e.g. in the co-leptogenesis models discussed in \cite{BB-PRL,EPJ-ST} 
which assumes that after inflation the O and M sectors are reheated asymmetrically, with $T \gg T'$, 
and masses of messenger ${\cal N}$ fermions between two sectors are larger than the 
reheating temperature. Nevertheless, the operators in \eqref{Ops}  
mediate scattering processes as $\ell_L \phi \to \bar{\ell}'_L\phi'$,  $\ell_L \phi \to \ell'_R \bar{\phi}'$ etc. 
which violate both $\rL$ and $\rL'$, and they are out-of-equilibrium. 
The invariance under ${\cal Z}^{LR}_2$ ($\ell_L \leftrightarrow \ell'_R$)
for the Yukawa couplings in \eqref{AApr}  
implies $Y'=Y^\ast$ and  ${\cal Y}'={\cal Y}^\ast$, in which case 
the  CP-violating factors in the above scattering processes are non-zero,  
and they induce ${\cal B}$ and ${\cal B}'$ of the opposite signs \cite{ADM,ADM-IJMP}. 
(Interestingly, this mechanism is ineffective in the case of ${\cal Z}_2$ symmetry 
yielding ${\cal Y}'={\cal Y}^\dagger$ since CP-violating factors appear to be vanishing \cite{ADM,ADM-IJMP}.)
Let us remark that this mechanism implies $\Omega_{b'} > \Omega_{b}$, 
which is related to the fact that M sector is colder and the produced $\rB'-\rL'$ 
suffers less damping  \cite{EPJ-ST}. Hence, it can naturally explain the 
observed cosmological fractions of  the baryons and DM, $\Omega_{b'}/\Omega_{b}\simeq 5$, 
which also makes clear who has eaten more oysters, the Walrus of the Carpenter. 
 
Hence, depending on the baryogenesis model and the type of discrete exchange symmetry, 
M sector can have positive or negative BA which sign should be also correlated to the 
chirality of M baryon in their weak interactions. 
In principle, two situations could be distinguished by the neutrino oscillations 
between two sectors. Namely, the ordinary SN explosions are believed to be accompanied by 
the short {\it neutronization} burst. 
Some part of them can oscillate into M neutrinos creating the deficit in the expected {\it neutrino} flux 
but this can be difficult to identify, mainly due to uncertainties in the SN core-collapse modelling.  
However, the neutronization bursts from the mirror SN explosions, depending on the case, 
could be observed (via the O and M neutrino oscillation) in terms of our 
{\it neutrinos}  or {\it antineutrinos}, without being accompanied by the optical SN. 
In particular, the observation of the {\it antineutronization} burst can be a smoking gun signal 
which could shade light on the mirror nature of DM and sterile neutrinos.

 \medskip 
 \noindent {\bf 6.} 
 Analogously to leptons, the mixed Lagrangian $\cL_\text{mix}$ can include the following 
D=9  operators  with quarks:
 \be{udd}
\frac{1}{\cM^5}\overline{(u' d^{\prime} d^{\prime})}_L(udd)_R ~+~ {\rm h.c.}  
\ee
where the parentheses contain the possible gauge invariant spin 1/2 chiral combinations 
of three O quarks and analogous combinations of three M quarks
(the gauge and Lorentz indices are omitted). 
These operators violate both $\rB$ and $\rB'$ by one unit  
but conserve $\widetilde{\rB}=\rB+\rB'$, 
and they induce $n-n'$ mass mixing \eqref{n-npr}   
with \cite{BB-nn'}:\footnote{We single out $n-n'$ mixing though generically operators \eqref{udd}  
can induce also other mixings as e.g. $\Lambda-\Lambda'$ between the O and M hyperons etc.}
\be{eps} 
\eps \simeq \frac{\Lambda_{\rm QCD}^6}{\cM^5} \simeq \left(\frac{10~{\rm TeV}}{\cM}\right)^5 
\times 10^{-15}~{\rm eV} 
\ee
 In an UV complete theory they can be induced via a seesaw like mechanism involving 
new heavy particles, as color-triplet scalars $S$ and $S'$ and a neutral Dirac fermion $F$, 
so that we have $\cM^5\sim M_S^4 M_{F}$ modulo the Yukawa coupling constants \cite{BB-nn',B-M}. 
Hence, for color scalars  at few TeV, the underlying theories can be testable 
at the LHC and future accelerators~\cite{B-M}.  
Interestingly, if ${\cal F}$ fermions are allowed to have also a small Majorana mass term, 
$\mu \ll M_{F}$, then the same seesaw mechanism would induce also 
$\Delta \rB=2$ operators $\sim (udd)_R^2$ (and their M counterparts) leading  
to the neutron-antineutron ($n-\ov{n}$) with mixing  $\eps_{n\bar{n}} = (\mu/M_F) \eps$  \cite{BB-nn',B-M}. 
Thus, the origin of both $n-n'$ and $n-\ov{n}$ oscillation phenomena 
can be related to the same new physics.  

However, $n-\bar n$ oscillation \cite{Kuzmin} is strongly restricted by 
the  direct experimental limit  
$\eps_{n\bar n} < 7.7 \times 10^{-24}$~eV,  while the nuclear stability bounds 
are yet stronger  yielding  $\eps_{n\bar n} < 2.5 \times 10^{-24}$~eV 
(for a review, see Ref.~\cite{Phillips}).  By this limits, $n-\ov{n}$ mixing 
cannot have a great effects on the neutron stars. The effective time of $n-\ov{n}$ 
conversion in the NS, which can be roughly estimated by replacing $\eps$ into $\eps_{n\bar n}$ 
in \eqref{taueps}, is larger than $10^{33}$~yr or so, while the energy production rate 
due to $n-\ov{n}$ annihilation can be no more than $10^{14}$~erg/s.

As for $n-n'$ oscillation, it has no restrictions from the nuclear stability \cite{BB-nn'}  
while the experimental limits \cite{Ban,Serebrov1,Altarev,Bodek,Serebrov2,ILL,Abel} 
still allow $n-n'$  mixing mass as large as $\eps \sim (10^{-15}\div 10^{-16})$~eV  
taking into account that the $n-n'$ oscillation probability can be suppressed by 
the environmental effects  
as e.g.  mirror magnetic fields which can be comparable to the normal magnetic 
field at the Earth \cite{More}.  As we have discussed in this paper, in this case 
the (anti)mirror NS can get  antimatter cores in their interior which can be rendered visible, 
via the baryon-antibaryon annihilation $\gamma$-rays, by accretion of 
ordinary matter from the ISM.  

We have discussed these effects in the minimal situation which assumes that 
$n-n'$ mixing occurs only due to mass mixing \eqref{n-npr}, $n-\bar{n}'$ mixing is absent, 
and $n$ and $n'$ are exactly degenerate in mass. 
However, the concept permits more variations which we briefly mention below: 

{\it Transitional magnetic moment.}  In difference from the $n-\ov{n}$ system 
where transitional magnetic moment between $n$ and $\ov{n}$ is forbidden by 
Lorentz invariance, non-diagonal magnetic moment $\mu_{nn'}$ (or dipole electric moment) 
is allowed between $n$ and $n'$ \cite{Arkady,MDPI} 
and they can be effectively induced in certain models of $n-n'$ mixing \cite{LHEP}. 
In this case the $n-n'$ transition time  will depend on the magnetic field 
in the NS, and it can be simply estimated by replacing $\eps$ into $\vert \mu_{nn'} B \vert$ 
in Eq.~\eqref{taueps}, or more concretely 
\be{eps-B}
\eps_{15} ~ \longrightarrow ~ \eps_{15}^B = \left(\frac{\mu_{nn'}  }{10^{-27}~\text{eV/G} } \right) 
\left(\frac{B}{10^{12}~G} \right) 
\ee 
Taking e.g. $\mu_{nn'} \sim 10^{-27}$~eV/G, which is 16 orders of magnitude smaller than 
the neutron magnetic moment itself, and making replacement \eqref{eps-B} in Eq.~\eqref{Nbar-dot}, 
we see that for a mirror magnetar ($B\sim 10^{15}$~G) the antimatter production rate 
will be $\sim 10^{40}/\text{s}$ while for an old recycled AMNS with $B \sim 10^8$~G it will be 
$\sim 10^{26}/\text{s}$. Thus, the former objects should be very bright 
while the latter can be practically invisible in annihilation $\gamma$-rays.  
Therefore, in this case the analysis similar to that of Ref.~\cite{Dupourque:2021poh}, 
would require a specific selection of the source samples which would take into 
account the distribution of magnetic field values in the NS. 

{\it $n-\bar{n}'$ mixing.}  For a simplicity, we have considered the situation 
with only $n-n'$ mixing \eqref{n-npr}, induced via effective D=9 operators \eqref{udd}, 
which conserves $\widetilde{\rB}=\rB+\rB'$.  
However, there can exist also $n-\bar{n}'$ mixing:  there is no fundamental reason to forbid it. 
However, the latter mixing, due to the SM structure, emerges from D=10 operators \cite{shortcut}, 
and the depending on the model parameters,  $n-\bar{n}'$ mixing mass $\eps_{n\bar{n}'}$ 
can be much smaller than $n-n'$ mixing mass $\eps$, but can be also comparable to it. 
In the latter case, with $\eps_{n\bar{n}'} \sim \eps_{nn'}$, both the MNS or AMNS could 
produce the baryon-antibaryon annihilation $\gamma$-rays, without `help' of the ordinary 
gas accretion.  Interestingly, in the presence of both $\eps_{n\bar{n}'}$ and $\eps_{nn'}$ 
with the comparable values is not conflict with the nuclear stability limits, while 
for the free neutron case it could allow to effectively induce $n-\bar{n}$ oscillation 
with pretty large rates provided that experimental conditions are properly tuned \cite{shortcut}. 

{\it $n-n'$ mass splitting.} We considered the minimal situation when $n$ and $n'$ 
have exactly the same masses in which case the experimental bounds 
\cite{Ban,Serebrov1,Altarev,Bodek,Serebrov2,ILL,Abel} imply $\eps < 10^{-15}$~eV or so. 
In this case the time of $n-n'$ transition \eqref{taueps}  is much larger than the Universe age, 
and thus it should be an ongoing process in the  existing  neutron stars (or M neutron stars). 
However, much larger values of $\eps$ are allowed by the experiment 
if  $n$ and $n'$ are not degenerate in mass. 
In particular, $n-n'$ oscillation e.g. with $\eps \sim 10^{-10}$~eV  or so can solve the neutron lifetime problem, 
 the $4\sigma$ discrepancy between the neutron lifetimes 
measured via the bottle and beam experiments,  provided that
$n$ and $n'$ have a mass splitting $m_{n'}-m_n\sim 100$~neV \cite{lifetime}.  
In fact, mass splitting will emerge  in models in which M parity is spontaneously broken \cite{BDM} 
but with a
rather small difference between the O and M Higgs VEVs $\langle \phi \rangle$ and $\langle \phi' \rangle$
\cite{Nussinov}.  
In this case $n-n'$ conversion time will be much smaller, $\tau_\eps \sim 10^6$~yr or so, 
so that the most of existing NS should be already transformed in maximally mixed stars 
with equal amounts and equal radii of the O and M components.
Hence, half of the AMNS in this case will be our antinuclear matter,  
and the $\gamma$-ray emission rate due to accretion will be given by Eq.~\eqref{J-ann}. 

Concluding, we have discussed a possibility of M world having a negative BA, 
so that the M neutron stars are the AMNS, and  $\bar{n'} - \bar{n}$ transition 
in their interior can create antimatter cores. The ordinary gas accreted from the ISM 
annihilating on the surface of these cores  give rise to $\gamma$-rays with the typical 
spectrum reducible to the baryon-antibaryon annihilation. 
  
 The alternative our mechanism 
 can be the existence of antimatter stars (antistars) \cite{Steigman:1976ev}.  
 The commonly accepted baryogenesis mechanisms fix the value as well as the sign 
 of the BA universally in the whole Universe. In addition, the observations 
 rule out the existence of significant amount of antimatter on scales ranging from 
 the solar system to galaxy and galaxy clusters,   
 and even at very large scales comparable to the present horizon \cite{Cohen,Steigman}. 
 However, more exotic baryogenesis mechanisms (for a review see \cite{Dolgov:1991fr}) 
 can in principle allow the existence of small domains at well-tempered scales 
 in which antimatter could survive in the form of anistars \cite{Khlopov-AS,Dolgov-AS}.  
 In particular,  the Affleck-Dine mechanism \cite{Affleck} can be extended by 
 the coupling of the Affleck-Dine $\rB$-charged scalar field to the inflaton 
 \cite{Dolgov-Silk}.  This modification, with properly tuned parameters, 
 can give rise to large baryon overdensities  at needed scales in which 
 the stars of specific pattern (or the baryon-dense objects (BDO) 
as they were named in Ref. \cite{Dolgov-BD})  can be formed. 
   In addition, in these overdensities the difference between 
the baryon and antibaryon amounts can be non-vanishing, and it could be positive as 
well as negative. Provided that part of the BDO consisting of antibaryons survive in the 
Milky Way (MW) halo up to present days,  they can be observed 
as the emitters of the $p\ov{p}$ annihilation $\gamma$-rays. 

In principle, the BDO and AMNS mechanisms can be distinguished by the spectral 
shape of the annihilation $\gamma$-rays. Namely, the proton annihilation on the surfaces 
of the BDO should produce $\gamma$-rays with typical spectrum peaked 
at 70 MeV or so \cite{Backenstoss:1983gu}. In the case of the AMNS, the spectral shape 
will be deformed by the surface redshift effect, by a factor $\exp[\phi]$, 
where $\phi=\phi(r)$ is the gravitational potential at the surface of antimatter core 
inside the AMNS. This will rescale down the spectral shape by $(15-30)\, \%$ 
depending on the AMNS mass, the EoS specifics and the radius of antimatter core $r<R$. 
In addition, one has to take into account the energy blueshift of the accreted protons:   
in fact, at the core surface they will be semi-relativistic, with the speeds 
nearly approaching the speed of light . 

In addition, let us recall that the AMNS can radiate substantial energy \eqref{E-dot} 
via the photons in the far UV/soft X-ray ranges (provided that the ratio \eqref{ratio} is 
much larger than 1) which can be an additional tracer for their 
identification. In addition, in Ref.~ \cite{Dupourque:2021poh} the sources possibly 
associated with pulsars were excluded from the possible antistar candidates. 
On the other hand, it is plausible that the AMNS are also observable as ordinary pulsars, 
if large ordinary magnetic fields are somehow developed in their antibaryon cores.  
This could be realized, for example, if along with $nn'$ mixing, there is also 
a kinetic mixing between the O and M photons \cite{Holdom} which effectively 
renders the mirror electrons and protons mini-charged 
(with tiny ordinary electric charges). The value of these electric mini-charges 
are severely restricted by the the cosmological \cite{Lepidi} 
and experimental \cite{Vigo} bounds.  Nevertheless, their existence can be effective.  
Since the antimatter core in the AMNS should consist of the heavy antinuclei and positrons, 
the AMNS rotation can induce circular electric currentsin its antimatter core 
by the drag mechanism~\cite{BDT} which can be sufficient 
for these cores to acquire significant magnetic field, as it was discussed in Ref.~\cite{BBMT}.  
Therefore, the AMNS could mimic ordinary pulsars, perhaps with some unusual 
properties. Having this in mind, maybe the pulsars should not be excluded 
from the candidate selection provided that their $\gamma$-emission has no 
high energy tail above 1 GeV or so.

\bigskip 

{\bf Acknowledgements} 
\medskip

\noindent
I thank Igor Tkachev for paying my attention to Ref. \cite{Dupourque:2021poh} and useful discussions. 
The work was supported in part by the research grant 
``The Dark Universe: A Synergic Multimessenger Approach" No. 2017X7X85K 
under the program PRIN 2017 funded by the Ministero dell'Istruzione, Universit\`a e della Ricerca (MIUR),  
and  in part by Shota Rustaveli National Science Foundation 
(SRNSF) of Georgia, grant DI-18-335/New Theoretical Models for Dark Matter Exploration. 


\begin{thebibliography}{99}


\bibitem{Alice}
  Z.~Berezhiani,
  Int.\ J.\ Mod.\ Phys.\ A {\bf 19}, 3775 (2004)  
  [arXiv:hep-ph/0312335];  
 Z.~Berezhiani, 
``Through the looking-glass: Alice's adventures in mirror world,''
 In  {\it From Fields to Strings, 
  Circumnavigating Theoretical Physics},  Eds. M. Shifman et al.,  vol. 3, pp. 2147-2195 
  [arXiv:hep-ph/0508233]  

\bibitem{Foot} 
  R.~Foot,
  Int.\ J.\ Mod.\ Phys.\ A {\bf 29}, 1430013 (2014)
  [arXiv:1401.3965 [astro-ph.CO]]

  
\bibitem{BB-nn'}
  Z.~Berezhiani and L.~Bento,
  Phys.\ Rev.\ Lett.\  {\bf 96}, 081801 (2006)   
 [arXiv:hep-ph/0507031] 
 
\bibitem{More} 
  Z.~Berezhiani,
  Eur.\ Phys.\ J.\ C {\bf 64}, 421 (2009) 
 [arXiv:0804.2088 [hep-ph]] 

\bibitem{Pokot} 
  Y.~N.~Pokotilovski,
  Phys.\ Lett.\ B {\bf 639}, 214 (2006)
  [arXiv:nucl-ex/0601017].

 \bibitem{HFIR}
 Z.~Berezhiani,  M.~Frost, Y.~Kamyshkov, B.~Rybolt and L.~Varriano,
  Phys.\ Rev.\ D {\bf 96},  no. 3, 035039 (2017)  
 [arXiv:1703.06735 [hep-ex]]  
  
\bibitem{UHECR} 
 Z.~Berezhiani and L.~Bento,
  Phys.\ Lett.\ B {\bf 635}, 253 (2006)  
 [arXiv:hep-ph/0602227] 

\bibitem{UHECR1} 
  Z.~Berezhiani and A.~Gazizov,
  Eur.\ Phys.\ J.\ C {\bf 72}, 2111 (2012)
 [arXiv:1109.3725 [astro-ph.HE]] 
 
\bibitem{Ban}
  G.~Ban {\it et al.},
 Phys.\ Rev.\ Lett.\  {\bf 99}, 161603 (2007)  
  [arXiv:0705.2336 [nucl-ex]]. 

\bibitem{Serebrov1}
  A.~Serebrov {\it et al.},
Phys.\ Lett.\  B {\bf 663}, 181 (2008)    
 [arXiv:0706.3600 [nucl-ex]].  

\bibitem{Altarev}
  I.~Altarev {\it et al.},
 Phys.\ Rev.\  D {\bf 80}, 032003 (2009)     
 [arXiv:0905.4208 [nucl-ex]].   

\bibitem{Bodek}
K.~Bodek {\it et al.},
 Nucl.\ Instrum.\ Meth. \  A {\bf 611}, 141 (2009). 
 
\bibitem{Serebrov2}
  A.~Serebrov {\it et al.},
 Nucl.\ Instrum.\ Meth. \  A {\bf 611}, 137 (2009)
 [arXiv:0809.4902 [nucl-ex]].  


\bibitem{ILL} 
  Z.~Berezhiani  {\it et al.}, 
  Eur.\ Phys.\ J.\ C {\bf 78},  717 (2018)
  [arXiv:1712.05761 [hep-ex]]
  
\bibitem{Abel} 
  C.~Abel {\it et al.},
  Phys.\ Lett.\ B {\bf 812}, 135993 (2021)
  [arXiv:2009.11046 [hep-ph]]

 \bibitem{Nesti}
  Z.~Berezhiani and F.~Nesti,
  Eur.\ Phys.\ J.\ C {\bf 72}, 1974 (2012) 
 [arXiv:1203.1035 [hep-ph]] 

\bibitem{Broussard} 
  L.~J.~Broussard {\it et al.},
  EPJ Web Conf.\  {\bf 219}, 07002 (2019)
  [arXiv:1912.08264 [physics.ins-det]].
  
\bibitem{ESS} 
  A.~Addazi {\it et al.},
J. Phys. G \textbf{48}, no.7, 070501 (2021)
[arXiv:2006.04907 [physics.ins-det]]

 
\bibitem{BBMT}
Z.~Berezhiani, R.~Biondi, M.~Mannarelli and F.~Tonelli,
arXiv:2012.15233 [astro-ph.HE]
 
\bibitem{INT} 
Z. Berezhiani, "Unusual effects in $n-n'$ conversion", 
talk at the Workshop INT-17-69W, Seattle, 23-27 Oct. 2017, 
\verb+http://www.int.washington.edu/talks/WorkShops/+ \verb+int_17_69W/People/Berezhiani_Z/Berezhiani3.pdf+ 


\bibitem{Massimo}
 M. Mannarelli, Z. Berezhiani, R. Biondi and F. Tonelli, ``nn$'$ conversion and neutron stars", 
 talk at NORDITA Workshop ``Particle Physics with Neutrons at the ESS", Stockholm, Sweden, 
 10--14 Dec. 2018, 
 \verb+https://indico.fysik.su.se/event/6570/timetable/+ \verb+#20181213+
 
\bibitem{LHEP} 
Z.~Berezhiani,
  LHEP {\bf 2}, no. 1, 118 (2019)
  [arXiv:1812.11089 [hep-ph]]

\bibitem{Nussinov-new} 
  I.~Goldman, R.~N.~Mohapatra and S.~Nussinov,
  Phys.\ Rev.\ D {\bf 100}, no. 12, 123021 (2019)
  [arXiv:1901.07077 [hep-ph]]

\bibitem{Ciancarella} 
  R.~Ciancarella, F.~Pannarale, A.~Addazi and A.~Marciano,
  Phys.\ Dark Univ.\  {\bf 32}, 100796 (2021)
  [arXiv:2010.12904 [astro-ph.HE]]
  
\bibitem{Makakin}
D.~McKeen, M.~Pospelov and N.~Raj,
[arXiv:2105.09951 [hep-ph]].
  
 \bibitem{ADM} 
 Z.~Berezhiani,
  arXiv:1602.08599 [astro-ph.CO] 

\bibitem{ADM-IJMP} 
  Z.~Berezhiani,
  Int.\ J.\ Mod.\ Phys.\ A {\bf 33}, no. 31, 1844034 (2018)

\bibitem{Dupourque:2021poh}
S.~Dupourqu\'e, L.~Tibaldo and P.~Von Ballmoos,
Phys. Rev. D \textbf{103}, no.8, 083016 (2021)
[arXiv:2103.10073 [astro-ph.HE]]

\bibitem{Fermi-LAT:2019yla}
S.~Abdollahi \textit{et al.} [Fermi-LAT],
Astrophys. J. Suppl. \textbf{247}, no.1, 33 (2020)
[arXiv:1902.10045 [astro-ph.HE]]

\bibitem{Backenstoss:1983gu}
G.~Backenstoss \textit{et al.}, 
Nucl. Phys. B \textbf{228}, 424-438 (1983)

\bibitem{Lattimer} 
  J.~M.~Lattimer,
  Ann.\ Rev.\ Nucl.\ Part.\ Sci.\  {\bf 62}, 485 (2012)
  [arXiv:1305.3510 [nucl-th]]; 
  I.~Vidana,
  Eur.\ Phys.\ J.\ Plus {\bf 133}, no. 10, 445 (2018)
  [arXiv:1805.00837 [nucl-th]] 

\bibitem{Sly} 
  F.~Douchin and P.~Haensel,
  Astron.\ Astrophys.\  {\bf 380}, 151 (2001)
  [arXiv:astro-ph/0111092].

\bibitem{Ozel} 
  F.~\"Ozel and P.~Freire,
  Ann.\ Rev.\ Astron.\ Astrophys.\  {\bf 54}, 401 (2016)
  [arXiv:1603.02698 [astro-ph.HE]]; 
  G.~F.~Burgio, H.~J.~Schulze, I.~Vidana and J.~B.~Wei,
  Symmetry {\bf 13}, no. 3, 400 (2021)

\bibitem{Bodmer} 
  A.~R.~Bodmer,
  Phys.\ Rev.\ D {\bf 4}, 1601 (1971); 
 E.~Witten, Phys.\ Rev.\ D {\bf 30}, 272 (1984)
 
\bibitem{BBDFL} 
  Z.~Berezhiani, I.~Bombaci, A.~Drago, F.~Frontera and A.~Lavagno,
  Astrophys.\ J.\  {\bf 586}, 1250 (2003)
  [arXiv:astro-ph/0209257] 

\bibitem{Schwab} 
  J.~Schwab, P.~Podsiadlowski and S.~Rappaport,
  Astrophys.\ J.\  {\bf 719}, 722 (2010)
  [arXiv:1006.4584 [astro-ph.HE]].

\bibitem{Alford}
M.~G.~Alford, A.~Schmitt, K.~Rajagopal and T.~Sch\"afer,
Rev. Mod. Phys. \textbf{80}, 1455-1515 (2008)
[arXiv:0709.4635 [hep-ph]]; 
R.~Anglani {\it at al.}, 
Rev. Mod. Phys. \textbf{86}, 509-561 (2014)
[arXiv:1302.4264 [hep-ph]]


\bibitem{Guillot} 
  S.~Guillot {\it et al.}, 
  Astrophys.\ J.\  {\bf 874}, no. 2, 175 (2019)
  [arXiv:1901.07998 [astro-ph.HE]]







\bibitem{Khlopov} 
  S.~I.~Blinnikov and M.~Y.~Khlopov,
  Sov.\ Astron.\  {\bf 27}, 371 (1983); 
  M.~Y.~Khlopov {\it et al.},
  Sov.\ Astron.\  {\bf 35}, 21 (1991); 
  H.~M.~Hodges,
  Phys.\ Rev.\ D {\bf 47}, 456 (1993)
  
  \bibitem{BDM}
  Z.~Berezhiani, A.~D.~Dolgov and R.~N.~Mohapatra,
  Phys.\ Lett.\ B {\bf 375}, 26 (1996)  
[arXiv:hep-ph/9511221];   
 Z..~Berezhiani,
 Acta Phys.\ Polon.\ B {\bf 27}, 1503 (1996) 
 [arXiv:hep-ph/9602326]; 
R.~N.~Mohapatra and V.~L.~Teplitz,
Astrophys. J. \textbf{478}, 29-38 (1997)
[arXiv:astro-ph/9603049]

  \bibitem{BCV}
  Z.~Berezhiani, D.~Comelli and F.~Villante,
  Phys.\ Lett.\ B {\bf 503}, 362 (2001)  
 [arXiv:hep-ph/0008105]   
 
\bibitem{IV}
  A.~Y.~Ignatiev and R.~R.~Volkas,
  Phys.\ Rev.\ D {\bf 68},  023518 (2003)  
[arXiv:hep-ph/0304260]

\bibitem{BCCV} 
  Z.~Berezhiani, P.~Ciarcelluti, D.~Comelli and F.~Villante,
  Int.\ J.\ Mod.\ Phys.\ D {\bf 14}, 107 (2005) 
 [arXiv:astro-ph/0312605]   

 \bibitem{BCCP} 
  Z.~Berezhiani, S.~Cassisi, P.~Ciarcelluti and A.~Pietrinferni,
  Astropart.\ Phys.\  {\bf 24}, 495 (2006)
  [astro-ph/0507153]

\bibitem{Roux}
J.~S.~Roux and J.~M.~Cline,
Phys. Rev. D \textbf{102}, no.6, 063518 (2020)
[arXiv:2001.11504 [astro-ph.CO]].

\bibitem{LIGO}
R.~Abbott \textit{et al.} [LIGO Scientific and Virgo],
Phys. Rev. X \textbf{11}, 021053 (2021)
[arXiv:2010.14527 [gr-qc]].

\bibitem{Merab}
R.~Beradze and M.~Gogberashvili,
Mon. Not. Roy. Astron. Soc. \textbf{487}, no.1, 650-652 (2019)
[arXiv:1902.05425 [gr-qc]]; 
MDPI Physics \textbf{1}, no.1, 67-75 (2019)
[arXiv:1905.02787 [gr-qc]] 

\bibitem{Merab-1}
  A.~Addazi and A.~Marcian\`o,
  Int.\ J.\ Mod.\ Phys.\ A {\bf 33}, no. 29, 1850167 (2018)
  [arXiv:1710.08822 [hep-ph]]; 
  R.~Beradze, M.~Gogberashvili and A.~S.~Sakharov,
  Phys.\ Lett.\ B {\bf 804}, 135402 (2020)
  [arXiv:1910.04567 [astro-ph.HE]]; 
  R.~Beradze and M.~Gogberashvili,
  Mon.\ Not.\ Roy.\ Astron.\ Soc.\  {\bf 503}, 2882 (2021)
  [arXiv:2101.12532 [astro-ph.CO]]



\bibitem{Sakh}
A.~D.~Sakharov,
Pisma Zh. Eksp. Teor. Fiz. \textbf{5}, 32 (1967)

\bibitem{KRS}
V.~A.~Kuzmin, V.~A.~Rubakov and M.~E.~Shaposhnikov,
Phys. Lett. B \textbf{155}, 36 (1985)

\bibitem{Mirror} 
  T.~D.~Lee and C.~N.~Yang,
  Phys.\ Rev.\  {\bf 104}, 254 (1956); 
  I.~Y.~Kobzarev, L.~B.~Okun and I.~Y.~Pomeranchuk,
  Sov.\ J.\ Nucl.\ Phys.\  {\bf 3}, no. 6,  837 (1966)   

\bibitem{FLV}
  R.~Foot, H.~Lew and R.~R.~Volkas,
 Phys.\ Lett. \ B {\bf 272}  67 (1991) 

\bibitem{Okun} 
L.~B.~Okun,
Phys.\ Usp.\  {\bf 50}, 380 (2007)
[arXiv:hep-ph/0606202]


\bibitem{Holdom} 
  B.~Holdom,
  Phys.\ Lett.\  B {\bf 166}, 196 (1986)

\bibitem{PLB}
Z.~Berezhiani,
Phys. Lett. B \textbf{417}, 287-296  (1998)
[arXiv:hep-ph/9609342]  

\bibitem{Belfatto}
B.~Belfatto and Z.~Berezhiani,
Eur. Phys. J. C \textbf{79}, no.3, 202 (2019)
[arXiv:1812.05414 [hep-ph]]; 
B.~Belfatto, R.~Beradze and Z.~Berezhiani,
Eur. Phys. J. C \textbf{80}, no.2, 149 (2020)
[arXiv:1906.02714 [hep-ph]]

\bibitem{ABK} 
  A.~Addazi, Z.~Berezhiani and Y.~Kamyshkov,
  Eur.\ Phys.\ J.\ C {\bf 77}, no. 5, 301 (2017)
[arXiv:1607.00348 [hep-ph]]

\bibitem{BGG}
Z.~Berezhiani, L.~Gianfagna and M.~Giannotti,
Phys. Lett. B \textbf{500}, 286-296 (2001)
[arXiv:hep-ph/0009290]

\bibitem{Cerulli} 
  R.~Cerulli {\it et al.}, 
  Eur.\ Phys.\ J.\ C {\bf 77},  no. 2, 83 (2017)
 [arXiv:1701.08590 [hep-ex]];  
  A.~Addazi {\it et al.}, 
  Eur.\ Phys.\ J.\ C {\bf 75}, no. 8,  400 (2015) 
  [arXiv:1507.04317 [hep-ex]].


\bibitem{Weinberg} 
  S.~Weinberg,
  Phys.\ Rev.\ Lett.\  {\bf 43}, 1566 (1979)

 \bibitem{ABS}
  E.~K.~Akhmedov, Z.~G.~Berezhiani and G.~Senjanovic,
  Phys.\ Rev.\ Lett.\  {\bf 69}, 3013 (1992) 
[arXiv:hep-ph/9205230];  
  R.~Foot, H.~Lew and R.~Volkas,
  Mod.\ Phys.\ Lett.\ A {\bf 7}, 2567 (1992);  
  R.~Foot and R.~Volkas,
  Phys.\ Rev.\ D {\bf 52}, 6595 (1995)  
 [arXiv:hep-ph/9505359]; 
  Z.~G.~Berezhiani and R.~N.~Mohapatra,
  Phys.\ Rev.\ D {\bf 52}, 6607 (1995)
 [arXiv:hep-ph/9505385]

 
 \bibitem{BB-PRL}
  L.~Bento and Z.~Berezhiani,
  Phys.\ Rev.\ Lett.\  {\bf 87}, 231304 (2001) 
[arXiv:hep-ph/0107281];   
L.~Bento and Z.~Berezhiani,
  Fortsch.\ Phys.\  {\bf 50}, 489 (2002);  
 %
 arXiv:hep-ph/0111116
%

\bibitem{EPJ-ST} 
 Z.~Berezhiani,
  Eur.\ Phys.\ J.\ ST {\bf 163}, 271 (2008)   

    \bibitem{B-M} 
  Z.~Berezhiani,
  Eur.\ Phys.\ J.\ C {\bf 76}, no. 12,  705 (2016) 
 [arXiv:1507.05478 [hep-ph]]   

\bibitem{Kuzmin} 
  V.~Kuzmin,
 Pisma Zh.\ Eksp.\ Teor.\ Fiz.\  {\bf 12}, 335 (1970);  
  R.~N.~Mohapatra and R.~E.~Marshak,
  Phys.\ Rev.\ Lett.\  {\bf 44}, 1316 (1980)  
  
  \bibitem{Phillips} 
 D.~G.~Phillips {\it et al.},
  Phys.\ Rept.\  {\bf 612}, 1 (2016)
[arXiv:1410.1100 [hep-ex]]; 
  K.~S.~Babu {\it et al.},
  arXiv:1310.8593 [hep-ex]; 
  arXiv:1311.5285 [hep-ph]

\bibitem{Arkady} 
  Z.~Berezhiani and A.~Vainshtein,
  Int.\ J.\ Mod.\ Phys.\ A {\bf 33}, no. 31, 1844016 (2018); 
  Phys.\ Lett.\ B {\bf 788}, 58 (2019)
  [arXiv:1809.00997 [hep-ph]]

\bibitem{MDPI} 
  Z.~Berezhiani, R.~Biondi, Y.~Kamyshkov and L.~Varriano,
  MDPI Physics {\bf 1}, no. 2, 271 (2019)
  [arXiv:1812.11141 [nucl-th]].

  \bibitem{shortcut} 
  Z.~Berezhiani,
  Eur.\ Phys.\ J.\ C {\bf 81}, no. 1, 33 (2021)
  [arXiv:2002.05609 [hep-ph]].


\bibitem{lifetime} 
  Z.~Berezhiani,
  Eur.\ Phys.\ J.\ C {\bf 79}, no. 6, 484 (2019)
  [arXiv:1807.07906 [hep-ph]].

\bibitem{Nussinov} 
  R.~N.~Mohapatra and S.~Nussinov,
  Phys.\ Lett.\ B {\bf 776}, 22 (2018) 
 [arXiv:1709.01637 [hep-ph]].


  
\bibitem{Steigman:1976ev}
G.~Steigman,
Ann. Rev. Astron. Astrophys. \textbf{14}, 339-372 (1976)

\bibitem{Cohen}
A.~G.~Cohen, A.~De Rujula and S.~L.~Glashow,
Astrophys. J. \textbf{495}, 539-549 (1998)
[arXiv:astro-ph/9707087 [astro-ph]]

\bibitem{Steigman}
G.~Steigman,
JCAP \textbf{10}, 001 (2008)
[arXiv:0808.1122 [astro-ph]]

\bibitem{Affleck}
I.~Affleck and M.~Dine,
Nucl. Phys. B \textbf{249}, 361-380 (1985)

\bibitem{Dolgov-Silk}
A.~Dolgov and J.~Silk,
Phys. Rev. D \textbf{47}, 4244-4255 (1993)

\bibitem{Dolgov:1991fr}
A.~D.~Dolgov,
Phys. Rept. \textbf{222}, 309-386 (1992)


\bibitem{Khlopov-AS}
M.~Y.~Khlopov {\it at al.}, 
Astropart. Phys. \textbf{12}, 367-372 (2000)
[arXiv:astro-ph/9810228 [astro-ph]]; 
M.~Y.~Khlopov, S.~G.~Rubin and A.~S.~Sakharov,
Phys. Rev. D \textbf{62}, 083505 (2000)
[arXiv:hep-ph/0003285 [hep-ph]]; 
K.~M.~Belotsky {\it at al.}, 
Phys. Atom. Nucl. \textbf{63}, 233-239 (2000)

\bibitem{Dolgov-AS}
C.~Bambi and A.~D.~Dolgov,
Nucl. Phys. B \textbf{784}, 132-150 (2007)
[arXiv:astro-ph/0702350 [astro-ph]];   
A.~D.~Dolgov, M.~Kawasaki and N.~Kevlishvili,
Nucl. Phys. B \textbf{807}, 229-250 (2009)
[arXiv:0806.2986 [hep-ph]]; 
A.~D.~Dolgov, S.~I.~Godunov, A.~S.~Rudenko and I.~I.~Tkachev,
JCAP \textbf{10}, 027 (2015)
[arXiv:1506.08671 [astro-ph.CO]] 

\bibitem{Dolgov-VV}
A.~D.~Dolgov, V.~A.~Novikov and M.~I.~Vysotsky,
JETP Lett. \textbf{98}, 519-522 (2013)
[arXiv:1309.2746 [hep-ph]]

\bibitem{Dolgov-BD}
S.~I.~Blinnikov, A.~D.~Dolgov and K.~A.~Postnov,
Phys. Rev. D \textbf{92}, no.2, 023516 (2015)
[arXiv:1409.5736 [astro-ph.HE]].

  
\bibitem{Lepidi} 
 Z.~Berezhiani and A.~Lepidi,
  Phys.\ Lett.\ B {\bf 681}, 276 (2009)
  [arXiv:0810.1317 [hep-ph]]

\bibitem{Vigo} 
  C.~Vigo {\it et al.}, 
  Phys.\ Rev.\ Lett.\  {\bf 124}, no. 10, 101803 (2020)
  [arXiv:1905.09128 [physics.atom-ph]].

\bibitem{BDT} 
  Z.~Berezhiani, A.~D.~Dolgov and I.~I.~Tkachev,
  Eur.\ Phys.\ J.\ C {\bf 73}, 2620 (2013)
  [arXiv:1307.6953 [astro-ph.CO]].

  
   
\end{thebibliography}
 \end{document}